\documentclass[sigconf, nonacm, natbib=false]{acmart} %

\acmConference[DAC '26]{Design Automation Conference}{July 2026}{Long Beach, CA}
\copyrightyear{2026}
\acmYear{2026}
\acmISBN{} %
\acmDOI{} %

\usepackage[
    backend=biber,
    style=numeric-comp,
    sorting=none,
    sortcites=true,
    url=true,
    eprint=true,
    giveninits=true,
    uniquename=true,
    minnames=1,
    maxnames=3
]{biblatex}
\AtEveryBibitem{%
    \clearname{editor}%
    \clearfield{series}%
    \clearfield{isbn}%
    \clearfield{issn}%
    \clearfield{day}%
    \clearfield{month}%
    \clearfield{place}%
    \clearlist{location}%
    \clearfield{eventtitle}%
    \ifentrytype{inproceedings}{%
        \clearlist{publisher}
    }{}%
    \ifentrytype{software}{%
    }{%
        \clearfield{url}%
        \clearfield{urlyear}%
    }%
}
\usepackage{siunitx}
\hypersetup{
    colorlinks=true,
    linkcolor=teal,
    citecolor=Maroon,
    urlcolor=Maroon
}

\newcommand{\condref}[1]{\hyperref[#1]{Condition~\ref*{#1}}}

\usepackage{booktabs}
\usepackage{multirow}
\usepackage{array}
\usepackage{xcolor}
\usepackage{graphicx}
\usepackage{subcaption}
\usepackage{tom}
\usetikzlibrary{calc, arrows.meta, trees, quotes, matrix, positioning}
\useyquantlanguage{groups}
\usepackage{tabularx}
\usepackage{environ}
\SetKwComment{Comment}{/* }{ */}

\makeatletter
\DeclareRobustCommand\rvdots{%
\vbox{%
\baselineskip4\p@\lineskiplimit\z@%
\kern-\p@%
\hbox{.}\hbox{.}\hbox{.}%
}%
}

\newcounter{problem}[section]
\renewcommand{\theproblem}{\arabic{problem}}
\newcommand{\problemtitle}[1]{\gdef\@problemtitle{#1}}
\newcommand{\probleminput}[1]{\gdef\@probleminput{#1}}
\newcommand{\problemquestion}[1]{\gdef\@problemquestion{#1}}
\NewEnviron{problem}{
  \refstepcounter{problem}%
  \problemtitle{}\probleminput{}\problemquestion{}%
  \BODY%
  \par\addvspace{.5\baselineskip}
  \noindent \textbf{Problem \theproblem: \@problemtitle}%
  \par\addvspace{.5\baselineskip}
  \noindent\begin{tabularx}{.5\textwidth}{@{\hspace{\parindent}} l X c}
    \textbf{Input:} & \@probleminput \\
    \textbf{Problem:} & \@problemquestion
  \end{tabularx}
  \par\addvspace{.5\baselineskip}
}
\makeatother

\crefname{problem}{Problem}{Problems}

\begin{document}

\title[Minimizing the Number of Code Switching Operation in Fault-Tolerant Quantum Circuits]{Minimizing the Number of Code Switching Operations\\in Fault-Tolerant Quantum Circuits}

\author{Erik Weilandt}
\email{erik.weilandt@tum.de}
\affiliation{%
  \department{Chair for Design Automation}
  \institution{Technical University of Munich}
  \country{Germany}
}

\author{Tom Peham}
\email{tom.peham@tum.de}
\affiliation{%
  \department{Chair for Design Automation}
  \institution{Technical University of Munich}
  \country{Germany}
}

\author{Robert Wille}
\email{robert.wille@tum.de}
\affiliation{%
  \department{Chair for Design Automation}
  \institution{Technical University of Munich}
  \country{}
}
\affiliation{%
  \institution{Munich Quantum Software Company}
  \country{Germany}
}

\begin{abstract}
    Fault-tolerant quantum computers rely on \emph{Quantum Error-Correcting Codes} (QECCs) to protect information from noise.
    However, no single error-correcting code supports a fully transversal and therefore fault-tolerant implementation of all gates required for universal quantum computation.
    Code switching addresses this limitation by moving quantum information between different codes that, together, support a universal gate set.
    Unfortunately, each switch is costly—adding time and space overhead and increasing the logical error rate.
    Minimizing the number of switching operations is, therefore, essential for quantum computations using code switching.
    In this work, we study the problem of minimizing the number of code switches required to run a given quantum circuit. 
    We show that this problem can be solved efficiently in polynomial time by reducing it to a minimum-cut instance on a graph derived from the circuit.
    Our formulation is flexible and can incorporate additional considerations, such as reducing depth overhead by preferring switches during idle periods or biasing the compilation to favor one code over another.
    To the best of our knowledge, this is the first automated approach for compiling and optimizing code-switching-based quantum computations at the logical level.
  \end{abstract}

\maketitle

\begin{figure*}[h!t]
    \centering
    \begin{subfigure}[b]{0.1\linewidth} %
        \centering
        \includegraphics[width=\linewidth]{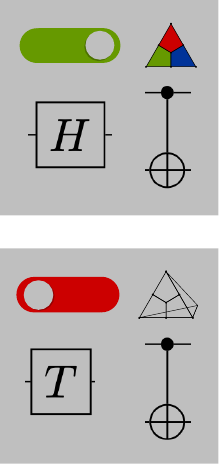}
        \caption{}
        \label{fig:code_switching_example_a}
    \end{subfigure}
    \hfill
    \begin{subfigure}[b]{0.44\linewidth}
        \centering
        \includegraphics[width=\linewidth]{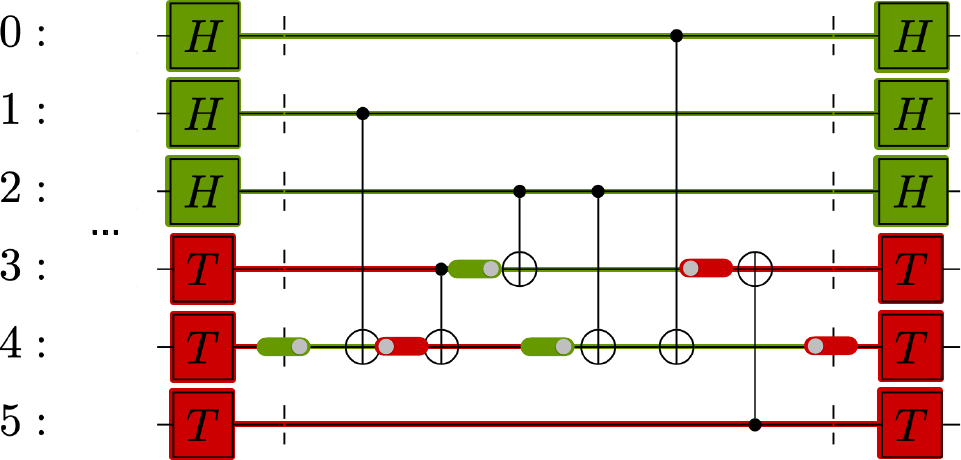}
        \caption{}
        \label{fig:code_switching_example_b}
    \end{subfigure}
    \hfill
    \begin{subfigure}[b]{0.44\linewidth}
        \centering
        \includegraphics[width=\linewidth]{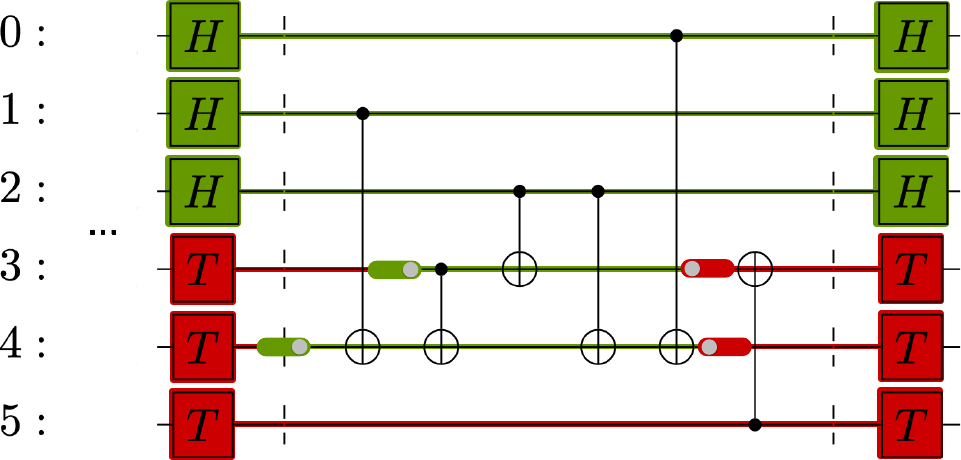}
        \caption{}
        \label{fig:code_switching_example_c}
    \end{subfigure}

    \caption{Minimal code switching: \textbf{(a)} Transversal gates of the 2D and 3D color codes. CNOT gates are transversal in both codes. \textbf{(b)} Circuit implementation requiring $6$ switching operations. \textbf{(c)} Minimal solution requiring only $4$ switches.}
    \label{fig:code_switching_example}
\end{figure*}

\section{Introduction}
\label{sec:intro}
Quantum computing~\cite{nielsenQuantumComputationQuantum2010} offers substantial speedups for certain classically hard tasks~\cite{dalzellQuantumAlgorithmsSurvey2025}.
However, the noisy nature of physical qubits requires the use of \emph{Quantum Error Correction Codes} (QECCs)~\cite{kitaevFaulttolerantQuantumComputation2003, shorFaulttolerantQuantumComputation1996, gottesmanTheoryFaulttolerantQuantum1998} to protect information during a computation.
Furthermore, computations have to be performed \emph{fault-tolerantly} to avoid small errors from affecting large parts of the encoded system.

One way to realize fault-tolerant computations is via \emph{transversal gates}, where logical gates are performed on an encoded system by acting on each qubit individually, thus preventing errors from spreading uncontrollably.
Unfortunately, no QECC with an encoded universal transversal gate set exists~\cite{eastinRestrictionsTransversalEncoded2009}, which means that the transversal gates of a code can only be used for restricted computations, which are not sufficient to solve problems of interest.

Two prominent proposals for overcoming these restrictions are:
\begin{itemize}
    \item \emph{Magic-state cultivation}~\cite{gidneyMagicStateCultivation2024, sahayFoldtransversalSurfaceCode2025, vakninEfficientMagicState2025} 
    and \emph{distillation} (MSD)~\cite{bravyiUniversalQuantumComputation2005, bravyiMagicstateDistillationLow2012, litinskiMagicStateDistillation2019}, 
    in which multiple noisy states are iteratively refined to produce fewer, higher-fidelity states that can then be used to implement logical gates via gate teleportation, and
    \item \emph{Code switching}~\cite{beverlandCostUniversalityComparative2021, buttFaultTolerantCodeSwitchingProtocols2024, heussenEfficientFaulttolerantCode2025, kubicaUniversalTransversalGates2015, bombinGaugeColorCodes2015}, 
    in which two QECCs with complementary transversal gate sets are used and logical qubits are switched between them as needed.
\end{itemize}

In the context of logical quantum circuit compilation where the task is to translate a given quantum circuit into instructions that can be executed for the target QECC and hardware platform, MSD-based computation in conjunction with lattice surgery~\cite{horsmanSurfaceCodeQuantum2012, litinskiGameSurfaceCodes2019, landahlQuantumComputingColorcode2014} has been a very active research field and many advances in compilation techniques have been made~\cite{tanSATScalpelLattice2024, herzogLatticeSurgeryCompilation2025, watkinsHighPerformanceCompiler2024, beverlandSurfaceCodeCompilation2022, laoMappingLatticeSurgerybased2018}.
However, despite some estimates showing that code switching may have lower overhead than magic state distillation in certain regimes~\cite{beverlandCostUniversalityComparative2021}, compilation for code switching-based fault-tolerant quantum computing has hardly been investigated.
While improvements have been made to reduce the overhead of the switching protocols themselves~\cite{buttFaultTolerantCodeSwitchingProtocols2024, heussenEfficientFaulttolerantCode2025}, to the best of our knowledge, no work exists that aims at optimizing code switching on the logical level. 

We formalize the \emph{minimal code switching problem}, which aims to identify the minimum number of code switches neededthroughout the execution of a circuit.
Because the gate sets supported by the different QECCs involved in the code switching scheme are not necessarily disjoint, i.e., there are some gates that can be executed in either code, it is not immediately clear when and where qubits should be switched to avoid unnecessary switching operations.
We show that this optimization problem can be solved in polynomial time by reducing it to the minimum cut problem~\cite{dantzigMaxflowMincutTheorem1957}.
Furthermore, we demonstrate that the min-cut formulation is highly flexible as it can be extended to incorporate one-way CNOTs between the QECCs, optimize circuit depth, and encode preferences for prioritizing one code over the other.

We implemented the proposed min-cut and heuristic approaches to the minimum code switching problem as part of the open source \emph{Munich Quantum Toolkit}~(MQT)~\cite{willeMQTHandbookSummary2024} (at \url{https://github.com/munich-quantum-toolkit/qecc}) and conducted a series of empirical evaluations on circuits up to $1024$ qubits and millions of gates.
The evaluations show that the minimum code switching problem can be solved in a moderate amount of time, even for circuits with up to $1024$ qubits and millions of gates.
We furthermore demonstrate that the proposed extensions to the min-cut approach allow the compiler to reduce circuit depth while preserving minimality with regard to the number of switching operations, and to trade a small number of additional switches for executing thousands more gates in a preferred code---an advantageous trade-off when the improved performance of that code outweighs the extra switching cost.

This work constitutes an important step in the compilation of code switching based fault-tolerant quantum computing. 
We hope that future developments in code switching compilation further help to evaluate when code switching might be preferable to MSD.

The rest of this work is structured as follows.
\Cref{sec:motivation} provides the necessary background on code switching and introduces the minimal code switching problem.
\Cref{sec:min-cut} shows how the minimal code switching problem can be reduced to min-cut and shows how the formulation can be extended to incorporate further features into the compilation process.
We evaluate the performance of the proposed algorithm in~\Cref{sec:eval} and demonstrate how the model can be tweaked to trade off minimality against other optimization criteria.
Finally, \Cref{sec:conclusion} concludes this paper.

\section{Background and Motivation}
\label{sec:motivation}
This section introduces and motivates the minimal code switching problem. 
To do this, we first revise the relevant background on code switching itself.

\subsection{Code Switching}
\label{sec:switching}

Different \emph{Quantum Error Correction Codes} (QECCs) support distinct sets of gates that can be implemented transversally. Transversal gates, which act on individual physical qubits of different logical code blocks, are inherently \mbox{fault-tolerant} as they do not spread errors uncontrollably through a quantum circuit. \emph{Code switching} has been proposed as a technique that employs multiple QECCs whose respective sets of transversal gates complement each other to achieve universality.
Logical qubits are dynamically transferred between these codes depending on which gate needs to be applied; in other words, the logical information is switched from one code to the other.
For the remainder of the paper, we consider the combination of a 2D~\cite{bombinTopologicalQuantumDistillation2006} and 3D color code~\cite{Bombin_2007} as a possible QECC pair for code switching.
\begin{example}
    2D color codes implement, among others, CNOT and Hadamard gates transversally. On the other hand, 3D color codes have CNOT and T gates in their transversal gate set. The union of both sets provides a universal gate set $\{H, T, \mathrm{CNOT}\}$ as sketched in~\Cref{fig:code_switching_example_a}.
\end{example}
Having identified two codes whose transversal gate sets complement each other, the next challenge is to leverage this combination in an actual computation. Code switching allows a computation to remain fault-tolerant throughout by temporarily encoding logical qubits in the code that is needed for the next gate to be executed transversally. In practice, this means the logical state must be transferred between codes whenever a gate outside the current code’s transversal set is needed.
\begin{example}
    Consider the input circuit shown in~\Cref{fig:code_switching_example_b}, which has been synthesized to only contain gates from the universal gate set. One can see that the single-qubit gates force the corresponding logical qubits into the respective code: Logical qubits on which a Hadamard gate acts have to be in the 2D color code (highlighted in green), and conversely, T gates enforce the 3D color code (highlighted in red). Assuming that a CNOT operation forces both participating qubits to be in the same code, we still have some degree of freedom to choose between the two codes. This is indicated by the switching symbols. So, for the first CNOT in~\Cref{fig:code_switching_example_b} we have the choice to either switch qubit 1 from green to red code or, as indicated, qubit 4 from red to green.
\end{example}

\subsection{Resulting Problem}
\label{sec:problem}
The process of switching between two codes introduces a non-trivial overhead into a computation, and different approaches to realize code switching have been proposed. 
It can be done by measuring certain gauge operators of a common subsystem code of the two codes in question~\cite{beverlandCostUniversalityComparative2021, buttFaultTolerantCodeSwitchingProtocols2024, kubicaUniversalTransversalGates2015, bombinGaugeColorCodes2015}, which requires fault-tolerant stabilizer measurements with flag circuits~\cite{chamberlandFlagFaulttolerantError2018}. Recently, a different technique has been proposed using doubled codes, which allows for teleporting the logical information into the other code using a one-way transversal CNOT~\cite{heussenEfficientFaulttolerantCode2025, sullivanCodeConversionQuantum2024}.
This requires fault-tolerant preparation of a high-quality logical ancillary system, which introduces additional overhead~\cite{forlivesiFlagOriginModular2025, zenQuantumCircuitDiscovery2025, pehamAutomatedSynthesisFaultTolerant2025}.

As a result, the frequency of switching operations directly influences both resource overheads and the overall logical error rate.
Consequently, a critical question arises at the logical level: \emph{how can we minimize the total number of switches required in a computation}? 
While much of the existing research has focused on the physical aspects of switching, the systematic optimization of switching patterns for given quantum circuits remains unexplored.

\begin{example}
    Consider~\Cref{fig:code_switching_example_c}, which realizes the same logical circuit as the one in~\Cref{fig:code_switching_example_b}. However, this circuit utilizes the degree of freedom  that CNOT gates can be realized in either code. 
    In fact, performing all CNOT gates except the last one in the 2D color code allows to reduce the number of switches from $6$ to $4$ (the actual minimum), reducing the overhead associated with switching.
\end{example}

Finding such optimal switching configurations is a \mbox{non-trivial} combinatorial challenge. Each possible assignment of logical qubits to different codes can affect the number and placement of required switches across multiple gates. As circuit size and connectivity increase, the space of possible switching patterns grows rapidly. Therefore, formally, we define:

\begin{problem}\label{prob:min-cs}
    \problemtitle{Minimal Code Switching Problem}
    \probleminput{Two quantum error correction codes with respective transversal gate sets $\Gamma_1$ and $\Gamma_2$ as well as a circuit composed of gates in  $\Gamma_1\cup\Gamma_2$.}
    \problemquestion{Determine the minimal number of switching operations and their locations required to execute the circuit using the available transversal gate sets.}
\end{problem}

Note that there are some circuit structures for which solving the minimal code switching problem is trivial. 
For example, the decomposition of single-qubit rotations into $H$- and $T$-gates requires potentially long sequences of alternating applications of $H$ and $T$~\cite{kitaevQuantumComputationsAlgorithms1997}.
In such decompositions, the minimum number of switches is obvious, as one has to switch after basically every gate.
The optimization potential lies in the part of the circuit that contains gates that can be performed in either code. 
In the following, we show how to exploit this optimization potential in polynomial time.

\section{Solving Minimal Switching with Min-Cut}
\label{sec:min-cut}

\begin{figure}
    \centering
    \includegraphics[width=.8\linewidth]{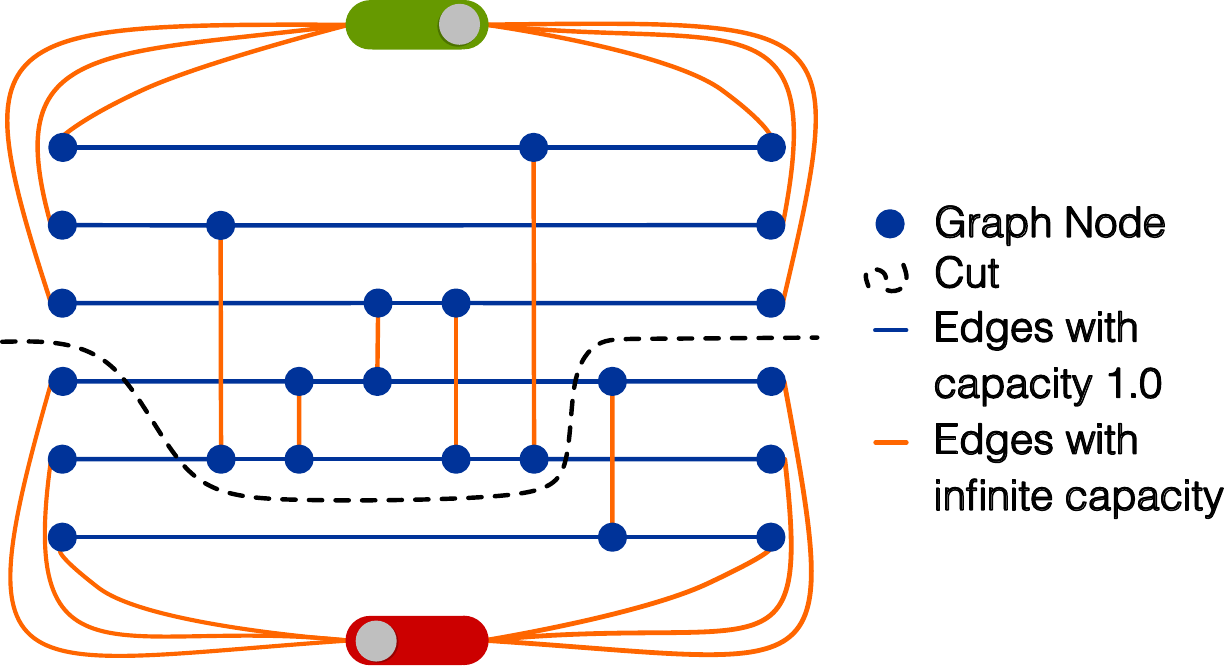}
        \caption{Network of the circuit in~\Cref{fig:code_switching_example}. Each qubit operation is represented as a graph node. The terminal nodes, depicted here as the green and red switches, represent the two different codes. The dashed line represents the minimum cut that separates the graph into two distinct subsets of nodes.}
        \label{fig:min-cut-graph}
\end{figure}

In this section, we demonstrate how to reduce the minimal code switching problem to solving an instance of the min-cut problem.
To lay the groundwork for this reduction, we will first review the necessary background on min-cut. Afterwards, the proposed solution, as well as possible extensions, are described in detail.
\subsection{Min-Cut}

A \emph{network} is a (directed) graph $G=(V,E,c)$, $E\subseteq V\times V$ with nodes $s, t \in V$ (often referrred to as \emph{source} and \emph{sink}) together with a \emph{capacity} function $c: E \rightarrow \R$.
An $s$-$t$ cut is a partition of $V$ into disjoint subsets $C=(S, T)$ with $S\sqcup T=V$ and a cut set $E_C=(S\times T) \cap E$ such that $s \in S$ and $t\in T$. 
In other words, removing $E_C$ from $E$ yields a graph $G'=(V, E\setminus E_C)$ such that there is no path from $s$ to $t$ in $G'$.
The cost $c(C)$ of a cut is the sum of the capacities of edges in the cut, i.e.,
\[c(C) = \sum_{e\in E_C} c(e).\]
There are usually many ways to construct an $s$-$t$ cut. 
A \emph{minimum cut} (or min-cut) is a cut with minimal costs.

The minimum $s$-$t$ cut problem is well studied especially due to its connection to maximum flow in flow networks~\cite{fordFlowsNetworks1962, tarjanAlgorithmsMaximumNetwork1986, dantzigMaxflowMincutTheorem1957}, and there are numerous algorithms that solve the min-cut problem efficiently in $O(|V|\cdot |E|^2)$~\cite{edmondsTheoreticalImprovementsAlgorithmic1972}, $O(|V|^2\cdot |E|)$~\cite{dinicAlgorithmSolutionProblem1970}, or $O(|V|^2\cdot \sqrt{|E|})$~\cite{goldbergNewApproachMaximumflow1988} time, and more efficient algorithms for specific classes of graphs have been developed as well~\cite{orlinMaxFlowsOnm2013}. 

\subsection{Reducing Minimal Code Switching to Min-Cut}
\label{sec:reduction}

To utilize min-cut solutions for the code switching problem, we first represent the input circuit as a graph.

Suppose we have a code-switching protocol between two codes supporting gate sets~$\Gamma_1$ and~$\Gamma_2$.  
Let the circuit be
\[
\mathcal{C} = g_m \circ g_{m-1} \circ \cdots \circ g_1,
\]
where each gate $g_i \in \Gamma = \Gamma_1 \cup \Gamma_2$ acts on some subset of $n$ qubits.

We construct a network $G = (V, E, c)$ as follows:

\begin{enumerate}
    \item \emph{Add source and sink nodes.}  
    Create two special nodes $s$ and $t$ representing the two codes that qubits can be assigned to.

    \item \emph{Add gate--qubit nodes.}  
    For every gate $g_i$ and every qubit $q_j$ it acts on, create a node 
    \[
    v(g_i, q_j) \in V.
    \]

    \item \emph{Add temporal edges.}  
    For each qubit $q$, connect the nodes corresponding to consecutive gates acting on $q$.  
    Add edges in both directions with capacity~$1$:
    \[
    (v(g_i, q), v(g_k, q)) \text{ and } (v(g_k, q), v(g_i, q)).
    \]

    \item \emph{Add gate edges.}  
    For every gate $g$ acting on qubits $Q(g) = \{q_1, \ldots, q_k\}$:
    \begin{itemize}
        \item If $g \in \Gamma_1 \setminus \Gamma_2$:  
        connect each $v(g, q) \in Q(g)$ to $s$ (and vice versa) with \emph{infinite} capacity.

        \item If $g \in \Gamma_2 \setminus \Gamma_1$:  
        connect each $v(g, q) \in Q(g)$ to $t$ (and vice versa) with \emph{infinite} capacity.

        \item In all cases, qubits acted on by the same gate must be in the same code.  
        Connect all nodes $\{v(g, q): q \in Q(g)\}$ to each other with \emph{infinite-capacity} edges.
    \end{itemize}
\end{enumerate}

\begin{example}
    Applying the procedure above to the circuit in~\Cref{fig:code_switching_example} results in the graph depicted in~\Cref{fig:min-cut-graph}.  
    Here, $\Gamma_1 = \{H, \mathrm{CNOT}\}$ and $\Gamma_2 = \{T, \mathrm{CNOT}\}$.
    We see that the nodes corresponding to $H$ gates are connected to the node representing the two-dimensional color code (green), and the nodes corresponding to $T$ gates are connected to the node representing the three-dimensional color code (red).
\end{example}

Given this graph-based representation of the circuit, the optimization task described in~\Cref{prob:min-cs} can be formulated as a graph partitioning problem: determining how to divide the graph into two subsets with the minimal number of cuts, which can be solved using a min-cut algorithm. The resulting subsets indicate which nodes belong to which terminal, or equivalently, which qubit operations should be executed in which code. Since cuts are only allowed along unit-capacity edges, i.e., before or after a gate, the resulting cut size directly corresponds to the number of required code switches, i.e., $c(C)=|E_C|$. The code switching locations are then determined by the cut temporal edges corresponding to positions in the circuit between gates.

\begin{example}
Consider the cut in~\Cref{fig:min-cut-graph} (indicated by a dashed line). All graph nodes below belong to the red code, and all nodes above to the green code, respectively. Each time the cut crosses a blue edge, it indicates a required cut of the edge, separating the two connected graph nodes, or in other words, a switching operation from one code to another between the two gate operations connected by the edge. This solution coincides with the solution shown in~\Cref{fig:code_switching_example_c} c).
\end{example}

\subsection{Extensions to the Min-Cut Model}
\label{sec:extensions}

The min-cut formulation introduced above can be extended to incorporate additional aspects of the compilation process, enabling further optimization in terms of both space and time overhead. 

For instance under specific conditions, CNOT operations can be implemented transversally even when the control and target qubits are encoded in different codes~\cite{heussenEfficientFaulttolerantCode2025,sullivanCodeConversionQuantum2024}. This property, however, is directional. In the 2D-3D color code scheme, it holds only when the control qubit is encoded in the 3D color code and the target qubit in the 2D color code\footnote{The direction depends on the exact definition of the 3D color code since the $X$- and $Z$-stabilizers are not symmetric, but it is always only one-way.}.

\begin{figure}[t]
    \centering
    \begin{subfigure}[b]{0.32\linewidth}
        \centering
        \includegraphics[width=\linewidth]{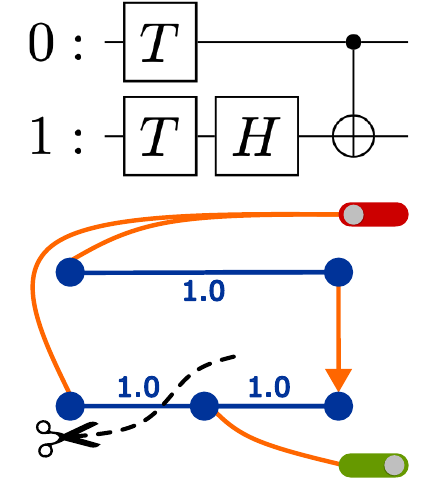}
        \caption{}
        \label{fig:optimization_graphs_one_way}
    \end{subfigure}
    \hfill
    \begin{subfigure}[b]{0.32\linewidth}
        \centering
        \includegraphics[width=\linewidth]{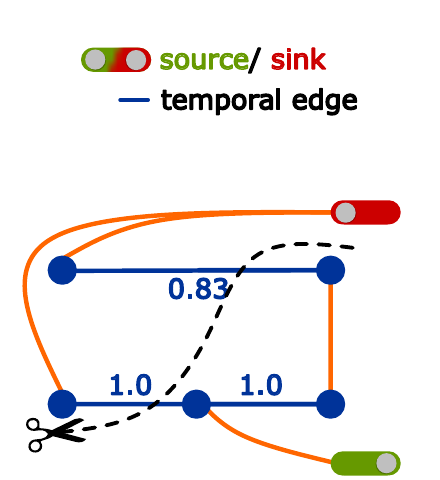}
        \caption{}
        \label{fig:optimization_graphs_depth}
    \end{subfigure}
    \begin{subfigure}[b]{0.32\linewidth}
        \centering
        \includegraphics[width=\linewidth]{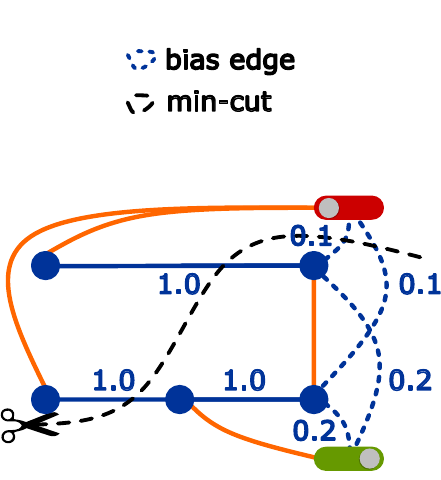}
        \caption{}
        \label{fig:optimization_graphs_bias}
    \end{subfigure}
    \hfill
    \caption{Modeling extensions to the min-cut formulation and resulting impact on the min-cut. (a) One-way transversal CNOT: cutting only a single edge separates source from sink.  (b) Prefer switching during qubit idling: a lower weight for idling edges pushes the min-cut algorithm to include these edges in the min-cut. (c) Code bias: adding additional bias edges and choosing capacities such that cutting bias edges of the 3D code results in lower overall costs, results in min-cuts that put more gate nodes on the 2D color code side of the cut.}
    \label{fig:optimization_graphs}
\end{figure}

To accurately model the one-way gates between different codes in the graph representation, we remove one of the two infinite-capacity edges in the encoding of CNOT gates. 
The cut set of an s-t cut removes edges so that no path from s to t exists. Paths from t to s are still allowed. If the edge is directed from control (source) to target (sink), then the cut must separate these terminals. Conversely, if the edge is directed from target to control, no path exists from source to sink, and no cut is needed. Thus, this modification does not introduce unnecessary cuts—it merely avoids adding cuts where they are not required, preserving the minimality of the solution.

\begin{example}
    \Cref{fig:optimization_graphs_one_way} shows the effect of this directed constraint. 
    The CNOT nodes are connected by a directed edge that goes from the control node to the target node. In this setup, the control node is associated with a 3D color code (sink node), while the target node is associated with a 2D color code (source node). 
    In this example, the source (green) can be separated from the sink (red) by cutting a single edge.
    Since the edge between the nodes involved in the CNOT is directed, no path exists from source to sink in the resulting graph.
    If the CNOT edge were bidirectional, at least one more edge would need to be cut.
    
\end{example}

As a result, circuits that exploit these one-way transversal interactions can achieve further reductions in switching overhead without requiring changes to the underlying optimization procedure.

Besides the minimal number of switching operations, the min-cut framework can also be extended to optimize additional circuit metrics. 
Consider, for example, the time overhead involved in realizing a code switching operation.
While code switching via transversal CNOTs has low overhead for the actual switching operations compared to switching via gauge measurements, it still requires fault-tolerant preparation of an ancilla, which must be completed before the switching takes place.
Due to this time overhead, it might be beneficial to perform switching operations during qubit idling.
We can achieve this behaviour by adding an \emph{idling bonus} to certain temporal edges. 
Specifically, temporal edges connecting consecutive nodes of an idling qubit are assigned a slightly reduced capacity. 

\begin{example}
\label{ex:optimization_depth}
    Consider the circuit in~\Cref{fig:optimization_graphs} in which qubit $0$ is idling for one single-qubit gate operation. 
    By assigning a lower capacity of $0.83$ to the edge connecting the node representing the $T$-gate and the control of the CNOT gate as shown in~\Cref{fig:optimization_graphs_depth}, the min-cut now has to cut this idling edge. 
    Performing the switch on the idling qubit~$0$ results in an overall shallower circuit as the switching can already start during the execution of the $H$ gate on qubit $1$.
\end{example}

When altering edge capacities in such a way, care must be taken so that the resulting min-cut does not lead to an increase in code switching operations compared to uniform temporal edge capacities.
If the capacities are too low, then cutting more low-capacity edges might have an overall lower cost than cutting fewer edges with higher capacity.
In the following, we briefly argue that the idling capacities can be chosen such that the cost of the resulting min-cut is still determined by the number of cut temporal edges.

\begin{figure}
    \centering
    \includegraphics[width=.8\linewidth]{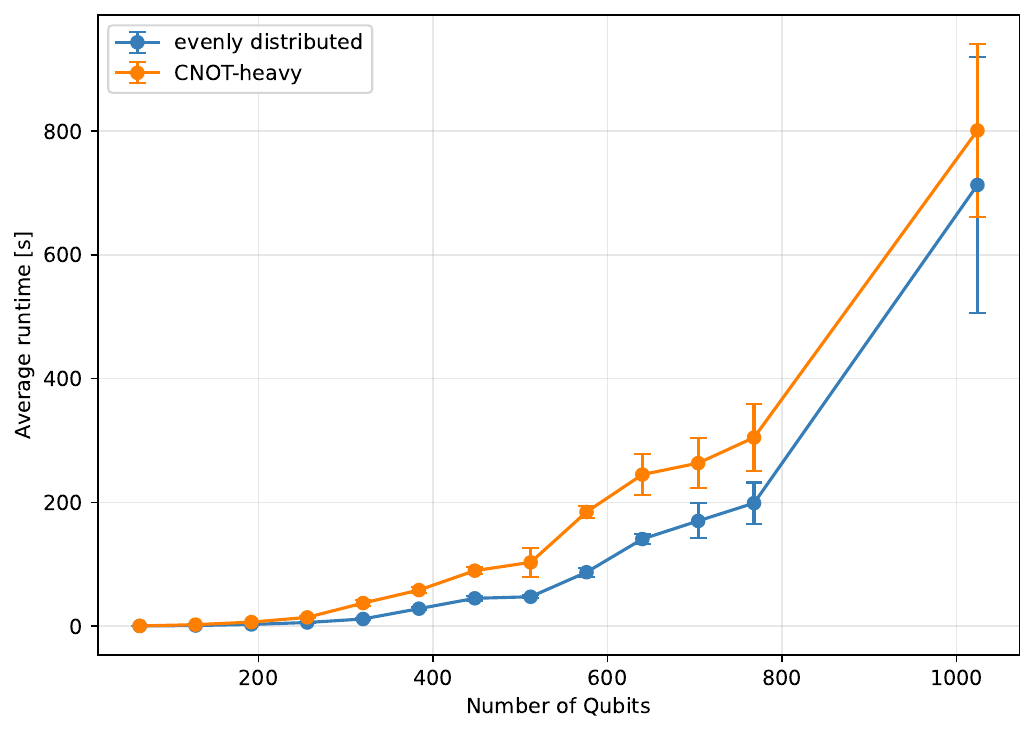}
        \caption{Average runtime of the proposed compilation methods across different circuit sizes and gate distributions.}
        \label{fig:results_average}
\end{figure}

If $E_\mathrm{temp} \subset E$ is the number of temporal edges, and $t_\mathrm{idle}(e)$ is the number of time steps the respective qubit idles between the gate nodes connected by $e$, then the capacity of a temporal edge $e$ is
\[c(e)=1-\frac{t_\mathrm{idle}(e)}{|E_\mathrm{temp}|(t_\mathrm{idle}(e)+1)}\]
The cost $c(C)$ of the cut $E_{C}$ is then given by
\begin{align*}
    c(E_{C}) &= \sum_{e\in E_{C}}c(e) = \sum_{e\in E_{C}}\left(1-\frac{t_\mathrm{idle}(e)}{|E_\mathrm{temp}|(t_\mathrm{idle}(e)+1)}\right) \\
    &= |E_{C}|-\left(\sum_{e\in E_{C}}\frac{t_\mathrm{idle}(e)}{|E_\mathrm{temp}|(t_\mathrm{idle}(e)+1)}\right).
\end{align*}
Now, since the size of the cut is at most as big as the set of all temporal edges, i.e., $|E_{C}| \leq |E_\mathrm{temp}|$, we can bound the second term as
\[0 \leq \sum_{e\in E_{C}}\frac{t_\mathrm{idle}(e)}{|E_\mathrm{temp}|(t_\mathrm{idle}(e)+1)} < \sum_{e\in E_{C}}\frac{1}{|E_\mathrm{temp}|} \leq 1,\]
and therefore
\[|E_C| \geq  c(E_C) > |E_C|-1.\]
This means that the cost of the cut is still determined by the number of temporal edges in the cut, and the influence of the idling bonus on the overall cost can never make up for cutting an additional edge. 
The idling bonus thus encodes the notion that performing switches during idle periods is “cheaper” in terms of circuit depth without sacrificing minimality with respect to the number of switches.

Another aspect to consider when determining suitable code switching locations is whether one of the codes involved is preferable to the other.
For example, the 3D color code has a higher code capacity threshold (for one type of error) than the 2D color code~\cite{kubicaThreeDimensionalColorCode2018, ohzekiAccuracyThresholdsTopological2009}.
It seems intuitive that the 3D color code would also exhibit a higher circuit-level threshold, due to the higher weight of the stabilizer generators that need to be measured. 
Since the minimal solution to the min-cut problem on the network is not necessarily unique, it may be preferable to choose a cut that performs more operations in the 2D code in this scenario.
It might even be beneficial to allow for slightly more switches if one can perform substantially more operations in the preferred code.

To embed this bias into the graph model, we introduce a new type of edge, referred to as \emph{bias edges}. 
Every node representing an operation that can be executed in either code is connected to both the source and the sink via a bias edge.
A bias edge connected to the source (corresponding to the preferable code) is assigned a slightly higher capacity than its counterpart connected to the sink. 
Because every such node is connected to both source and sink, any min-cut in this modified graph needs to cut one of the bias edges. By assigning a higher capacity to the preferred code, the min-cut is therefore skewed towards cutting bias edges connected to the node of the less desirable code and, therefore, performing more operations in the preferred code. 

\begin{example}
    Consider the example circuit in~\Cref{fig:optimization_graphs} again.
    \Cref{fig:optimization_graphs_bias} shows the corresponding network of this circuit using bias edges. 
    Bias edges connected to the 2D color code have a bias of $0.2$, while bias edges connected to the 3D color code have a bias of $0.1$.
    The min-cut, therefore, cuts the edge on qubit $0$ as well as the bias edge connecting the CNOT nodes to the 3D color code node.
    This corresponds to switching qubit $0$ into the $2D$ color code.
\end{example}

By picking capacities for these bias edges that are only a small fraction of those of temporal edges, it can be ensured that they influence the optimization without dominating it. 
Importantly, the algorithm must still cut temporal edges to produce a valid solution; it cannot minimize the cost solely by cutting bias edges, which means that a switching scheme can still be extracted from the cut in this modified network. 
However, depending on the capacity of the bias edges, a min-cut in this modified network might correspond to more switching operations if a significant amount of bias edges can be cut instead.

\section{Evaluations}
\label{sec:eval}
\begin{figure}[]
    \centering
    \includegraphics[width=.8\linewidth]{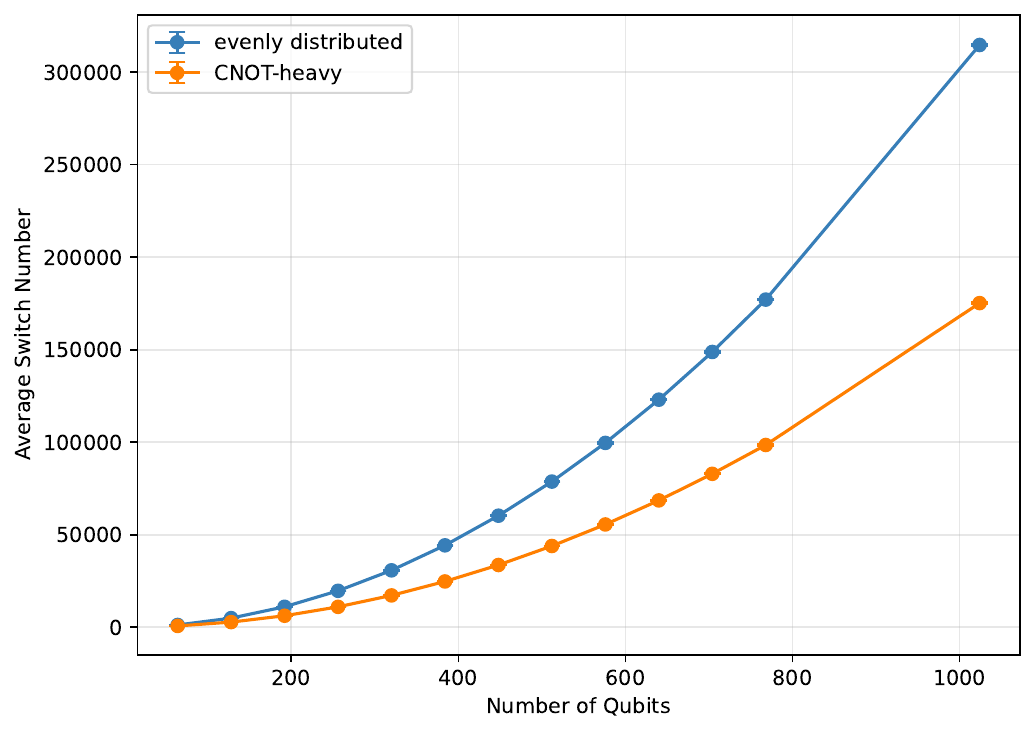}
        \caption{Average number of minimal switching operations across different circuit sizes and gate distributions.}
        \label{fig:avg_switch_number}
\end{figure}
We implemented the proposed min-cut-based approach to solving the minimal code switching problem in Python within an existing open-source quantum compilation framework.
For finding the min-cut, we used the open-source graph library NetworkX~\cite{Newman_2003}.

\subsection{Experimental Setup}

To evaluate the proposed approach across a broad spectrum of benchmark circuits, we generate quantum circuits using gates from the set $\{H,T,\mathrm{CNOT}\}$ as well as the identity gate, for varying circuit sizes.
We assume a switching scheme using 2D and 3D color codes with transversal gate sets $\{H, CNOT\}$ and $\{T, CNOT\}$, respectively.
Because this setting allows for one-way transversal CNOTs, they are modeled in the respective graphs by default.
When generating the circuits, each gate has a predefined probability of being applied to a qubit at each step, with the constraint that, except for the identity, the same gate cannot be applied twice in a row.
We construct two circuit classes: \emph{evenly distributed} circuits, containing $15\%$ each of $H$, $T$, and $\mathrm{CNOT}$ gates, and \emph{CNOT-heavy} circuits, containing $10\%$ $H$ and $T$ gates, and $30\%$ $\mathrm{CNOT}$ gates.

For performance benchmarking, we evaluated circuits with qubit counts ranging from $n = 64$ to $n = 1024$, sampling $2n$ gates per qubit. For each qubit count, we generate $100$ circuits with an even gate distribution and $100$ CNOT-heavy circuits. To further evaluate extensions to our model (e.g., bias edges or idling bonuses), an additional $1000$ circuits were generated with $n = 128$ qubits and $2n$ gates per qubit sampled from the even gate distribution.

All experiments were conducted on an Ubuntu 20.04 system equipped with an AMD Ryzen Threadripper PRO 5955WX CPU (16 cores, 32 threads, 4.0–4.5 GHz) and 128 GB of DDR4 RAM.

\subsection{Performance Evaluation}

\Cref{fig:results_average} shows the average runtime for computing the minimal number of code switching operations on the generated circuits.  Even for circuits with $1024$ qubits and approximately a million gates, the algorithm still only takes about \SI{800}{\second}. 
This can potentially be further optimized by using a more efficient implementation of the min-cut algorithm.
Moreover, it can be observed that the runtime is influenced by the circuit structure. CNOT-heavy circuits---involving more gates that can be executed in either code---exhibit longer runtimes than circuits containing more $H$ and $T$ gates, which restrict code choices and reduce combinatorial complexity.

\Cref{fig:avg_switch_number} shows the average number of minimal switches required for the considered benchmark circuits.
Evenly distributed circuits need roughly twice as many code switching operations as CNOT-heavy circuits. This is because a larger fraction of consecutive single-qubit gates ($H$ and $T$) results in more unavoidable switching locations, limiting optimization potential.

\subsection{Idling Qubits and Code Bias}

\begin{table}[t]
    \centering
    \caption{Relative reduction in circuit depth achieved when incorporating idling time into the capacity of temporal edges compared to uniform capacities.}
    \begin{tabular}{ccccc}
        \toprule
        \textbf{\#Qubits} & \textbf{Relative Saving (\%)} & \textbf{Relative Std (\%)} \\
        \midrule
        64  & 5.25  & 3.92 \\
        128 & 4.79  & 2.02 \\
        256 & 5.56  & 1.55 \\
        512 & 5.41  & 0.97 \\
        \bottomrule
    \end{tabular}
    \label{tab:depth_reduction}
\end{table}

In a second series of evaluations, we explore the potential for reducing circuit depth by incorporating an idle bonus on temporal edges representing idling qubits.
To evaluate the impact of this extension on the considered circuits, we  compare the depth of circuits with and without idling bonus as discussed in~\Cref{sec:extensions}.
The specific time requirement for code switching ultimately depends on the specific code switching scheme in use, as well as other factors such as the specific hardware platform.
In our evaluations, we assume that each switch takes longer than a logical single-qubit gate but no longer than two consecutive logical single-qubit operations in our simulations. 

The average circuit depth reduction achieved through idle-based optimization under this assumption on circuits using an even distribution of gates is shown in~\Cref{tab:depth_reduction}. 
Across different numbers of qubits, the depth saving is about $5\%$ even under this very optimistic assumption.
If switching takes longer, incorporating qubit idling into the choice of switching locations will have an even greater impact.
Moreover, the results confirm that including this aspect in the proposed approach does not violate the minimality of the solution.

\begin{figure}
    \centering
    \includegraphics[width=.8\linewidth]{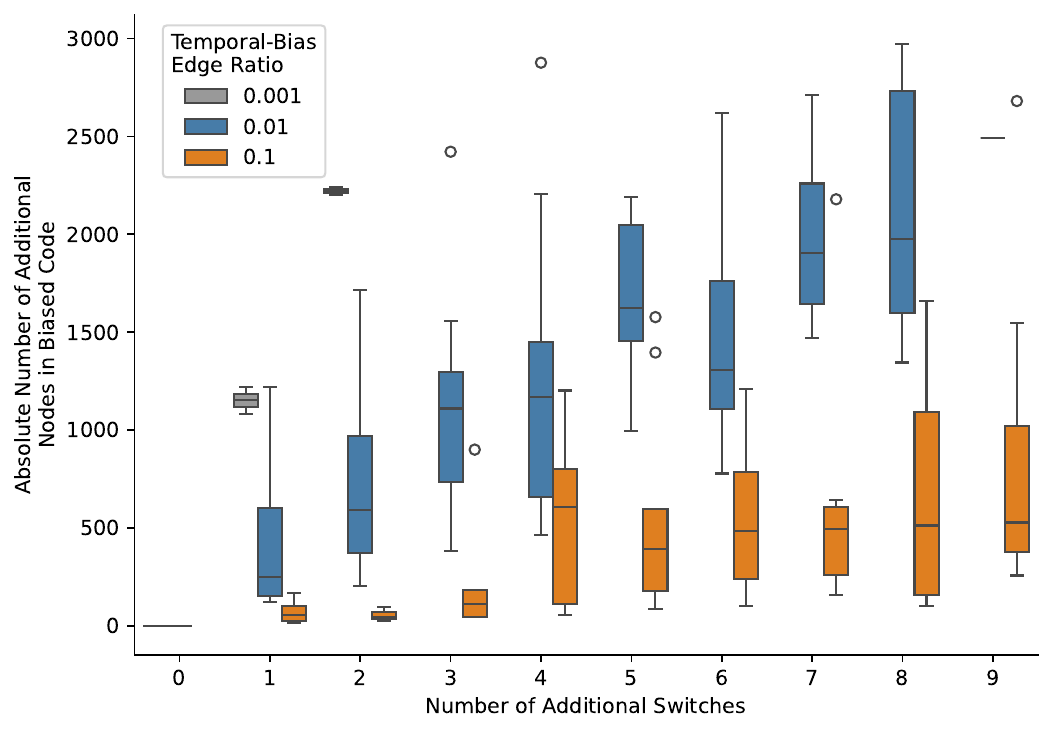}
        \caption{Boxplot illustrating the number of additional nodes in the 2D color code when allowing for extra switching operations compared to the optimal solution.}
        \label{fig:bias-saving}
\end{figure}

Finally, we examine how varying the ratio between the capacities of temporal and bias edges affects the trade-off between preserving optimality and reducing the number of nodes assigned to the 3D color code.
\Cref{fig:bias-saving} shows how additional switching operations (x-axis) influence the number of extra nodes in the biased code (y-axis). 
We compute the min-cut for circuits using ratios $0.1$, $0.01$, and $0.001$.
The capacity of the bias edges for the 2D color code is chosen as twice as large as the bias for the 3D color code. 
The results demonstrate that a single additional switching operation can lead to several hundred nodes shifting from the 3D to the 2D color code.
Smaller ratios result in  a larger trade-off.
This is expected, as the cost of a single additional temporal edge in the cut must be compensated for by more biased edges. 
If the ratio is $0.01$, one cut of a temporal edge results in at least $100$ more cuts on bias edges.

Interestingly, no change in the number of operations in the 2D color code can be observed, i.e., the min-cut without modeling code bias already maximizes the number of CNOTs performed in the 2D code.
This is because the min-cut implementation we used puts the majority of nodes on the source side of the cut by default, and we represent the 2D color code as the source node.

\section{Conclusion}

In this work, we investigated the minimal code switching problem for quantum circuits and demonstrated how it can be solved efficiently by reducing it to the minimum $s$-$t$-cut problem.
We also showed how additional circuit metrics can be incorporated into the min-cut framework and optimized simultaneously.
Furthermore, we implemented and evaluated the proposed solution on a large number of benchmarks to demonstrate its scalability to circuits with up to $1024$ qubits and approximately a million gates, as well as how extensions such as code bias and idle-aware switching enable meaningful trade-offs between minimal switching, circuit depth, and code preferences.

The runtime efficiency of our approach makes it a good candidate for use as a cost function in quantum circuit optimization.
Given a choice of different circuits implementing the same unitary, one can quickly evaluate which one requires the fewest switches, or which one realizes a preferable compromise between switching overhead and other circuit-level metrics.

This work constitutes a significant advancement in the compilation of code-switching-based fault-tolerant quantum computing.
We view our contributions as complementary to advancements in decoding and physical-level implementations of code switching, and we expect that the proposed techniques will help in evaluating the practicality of code switching as a candidate for universal fault-tolerant quantum computation.

\label{sec:conclusion}

\section*{Acknowledgements}
\small
The authors would like to thank Sascha Heußen for insightful discussions and comments.

The authors acknowledge funding from the European Research Council (ERC) under the European Union’s Horizon 2020 research and innovation program (grant agreement No.\ 101001318) and Millenion (grant agreement No.\ 101114305). 
This work was part of the Munich Quantum Valley, which is supported by the Bavarian state government with funds from the Hightech Agenda Bayern Plus.
This work was funded by the Deutsche Forschungsgemeinschaft (DFG, German Research Foundation, No. 563402549).

\balance
{\printbibliography}
\end{document}